\documentclass{PoS}
\usepackage{amsmath,amssymb,amsfonts}
\usepackage{graphicx,epsfig,url}
\usepackage{float}
\usepackage{amsmath}
\usepackage{amsfonts}
\usepackage{amssymb}
\usepackage{graphicx}

\def\({\left(}
\def\){\right)}
\def\f{\frac}
\def\be{\begin{equation}}
\def\ee{\end{equation}}

\def\de{\partial}
\def\demub{\de_{\mu}}
\def\denub{\de_{\nu}}

\input{tcilatex}
\usepackage{cite}

\title{Composite vectors in Higgless models at the LHC}

\ShortTitle{Composite vectors in Higgless models at the LHC}

\author{\speaker{Antonio Enrique C\'arcamo Hern\'andez}%
         \thanks{A footnote may follow.}\\
        Scuola Normale Superiore Di Pisa and  INFN, Sezione di Pisa, Pisa, Italy\\
        E-mail: \email{antonio.carcamo@sns.it}}

\abstract{In the context of a strongly coupled Electroweak Symmetry Breaking, composite triplet of heavy vectors degenerate in mass belonging to the $SU(2)_{L+R}$ adjoint representation may arise from a new strong interaction invariant under the global $SU(2)_L\times SU(2)_R$ symmetry, which is spontaneously broken down to $SU(2)_{L+R}$. Assuming that the interactions among these heavy vector states and with the Standard Model gauge bosons are described by a $SU(2)_L\times SU(2)_R/SU(2)_{L+R}$ Effective Chiral Lagrangian, the heavy vector pair production at the LHC by Vector Boson Fusion and Drell-Yan annihilation is studied in this framework. The expected rates of same sign di-lepton and tri-lepton events from the decay of the composite vectors into Standard Model gauge bosons are computed. }

\FullConference{XVIII International Workshop on Deep-Inelastic Scattering and Related Subjects\\
April 19 -23, 2010\\
Convitto della Calza, Firenze, Italy}

\begin{document}

\section{Introduction}
It is known that there are two pictures used to explain the mechanism of Electroweak Symmetry Breaking, that is, weakly and strongly coupled. The LHC is going to determine which of these pictures is realized by nature. A weakly coupled dynamics describing the mechanism of the Electroweak Symmetry Breaking (EWSB) is provided by the Standard Model and its Supersymmetric extensions. In the Standard Model, the existence of one Higgs doublet is assumed in order to explain the generation of the masses of all the fermions and bosons. In addition to the 3 eaten up Goldstone bosons, the Higgs doublet contains one physical neutral scalar particle, called the Higgs boson, which is crucial for keeping under control unitarity in the elastic and inelastic channels of the gauge boson scattering and which allows us to extrapolate a weakly coupled model up to the Planck scale. A light Higgs boson can also successfully account for the Electroweak Precision Tests (EWPT).
 In spite of the very good agreement of the Standard Model predictions with experimental data, the Higgs boson has not been found and then, the mechanism of Electroweak Symmetry Breaking responsible for the generation of the masses of the fermions and bosons remains to be explained. Besides that, the Standard Model has the so called hierarchy problem, which is the strong sensitivity of the Higgs mass to very short distance scales. These problems provide plausible motivations for considering strongly coupled theories of EWSB, which become non-perturbative above the Fermi scale and the breaking is achieved through some condensate. In the most general framework of a strongly interacting dynamics for EWSB, one can have composite resonances which could be spin-0, spin-1/2 and spin-1 states. These composite particles are bound states of more fundamental constituents which are held together by a new strong interaction. Here the case in which the strong dynamics responsible for EWSB gives rise to a triplet of composite vectors belonging to the adjoint representation of the custodial symmetry group $SU(2)_{L+R}$ is considered. These composite vectors states are very important in keeping under control the asymptotic behaviour of the longitudinal SM gauge boson scattering amplitude up to a cutoff $\Lambda \simeq 3$ TeV and under suitable conditions can account for Electroweak Precision Tests \cite{Barbieri:2008}. This work is devoted to the study of the heavy vectors pair production at the LHC by Vector Boson Fusion and Drell-Yan annihilation. 
\section{Effective Chiral Lagrangian with massive spin one fields.}
In the framework of a strongly interacting dynamics for EWSB, the interactions among the composite triplet of heavy vectors and the SM particles can be described by the following model independent $SU(2)_L\times SU(2)_R/SU(2)_{L+R}$ Chiral Lagrangian invariant under parity \cite{Barbieri:2010}:
\begin{eqnarray}
\mathcal{L} &=&\frac{v^{2}}{4}\left\langle D_{\mu }U\left( D^{\mu
}U\right) ^{\dag }\right\rangle -\frac{1}{2g^{2}}\left\langle W_{\mu \nu
}W^{\mu \nu }\right\rangle -\frac{1}{2g^{\prime 2}}\left\langle B_{\mu \nu
}B^{\mu \nu }\right\rangle-\frac{1}{4}\left\langle \hat{V}^{\mu \nu }\hat{V}_{\mu \nu }\right\rangle
+\frac{M_{V}^{2}}{2}\left\langle {V}^{\mu }{V}_{\mu }\right\rangle\notag \\
&&-\frac{%
ig_{V}}{2\sqrt{2}}\left\langle \hat{V}^{\mu \nu }[u_{\mu },u_{\nu
}]\right\rangle-\frac{f_{V}}{2\sqrt{2}}\left\langle \hat{V}^{\mu \nu }(u{W}_{\mu \nu
}u^{\dag }+u^{\dag }{B}_{\mu \nu }u)\right\rangle +\frac{ig_{K}}{2\sqrt{2}}%
\left\langle \hat{V}_{\mu \nu }{V}^{\mu }{V}^{\nu }\right\rangle \notag \\
&&+g_{1}\left\langle {V}_{\mu }{V}^{\mu }u^{\alpha }u_{\alpha }\right\rangle
+g_{2}\left\langle {V}_{\mu }u^{\alpha }{V}^{\mu }u_{\alpha }\right\rangle
+g_{3}\left\langle {V}_{\mu }{V}_{\nu }[u^{\mu },u^{\nu }]\right\rangle+g_{4}\left\langle {V}_{\mu }{V}_{\nu }\{u^{\mu },u^{\nu }\}\right\rangle 
\notag \\
&&
+g_{5}\left\langle {V}_{\mu }\left( u^{\mu }{V}_{\nu }u^{\nu }+u^{\nu }{V}%
_{\nu }u^{\mu }\right) \right\rangle+ig_{6}\left\langle {V}_{\mu }{V}_{\nu }(u{W}^{\mu \nu }u^{\dag }+u^{\dag }%
{B}^{\mu \nu }u)\right\rangle
\end{eqnarray}
where $f_V=\frac{F_V}{M_V}$, $g_V=\frac{G_V}{M_V}$, $g_K$, $g_i$ with $i=1,2,\cdots,6$ are dimensionless parameters and the following relations are satisfied:
\begin{equation}%
\begin{array}
[c]{l}%
U\left(  x\right)  =e^{i\hat{\pi}\left(  x\right)  /v}\,,\qquad\hat{\pi
}\left(  x\right)  =\tau^{a}\pi^{a}=\left(
\begin{array}
[c]{cc}%
\pi^{0} & \sqrt{2}\pi^{+}\\
\sqrt{2}\pi^{-} & -\pi^{0}%
\end{array}
\right),\\
D_{\mu}U=\partial_{\mu}U-i{B}_{\mu}U+iU{W}_{\mu}\,,\qquad{W}_{\mu}=\frac
{g}{{2}}\tau^{a}W_{\mu}^{a}\,,\qquad{B}_{\mu}=\frac{g^{\prime}}{{2}}\tau
^{3}B_{\mu}^{0}\,,\\
\displaystyle W_{\mu\nu}=\demub W_{\nu}-\denub W_{\mu}-i[W_{\mu},W_{\nu}]\,,\qquad\displaystyle B_{\mu\nu}=\demub B_{\nu}-\denub B_{\mu}\,
\end{array}
\label{eq2}%
\end{equation}
The field strength tensor $\hat{V}_{\mu\nu}=\nabla_{\mu}V_{\nu}-\nabla_{\nu}V_{\mu}$ is written in terms of the $SU(2)_{L}\times SU(2)_{R}$ covariant derivative
\be
\nabla_{\mu}V_{\nu}=\demub V_{\nu}+[\Gamma_{\mu},V_{\nu}]\,,\qquad V_{\mu}=\f{1}{\sqrt{2}}V_{\mu}^{a}\tau^{a}
\ee
with the connection $\Gamma_{\mu}$ given by
\be
\Gamma_{\mu}=\f{1}{2}\Big[u^{\dag}\(\demub-iB_{\mu}\)u+u\(\demub-iW_{\mu}\)u^{\dag}\Big]\,,\qquad u\equiv \sqrt{U}\,,\qquad \Gamma_{\mu}^{\dag}=-\Gamma_{\mu}\,.
\ee
and $u_{\mu}=u^{\dag}_{\mu}=iu^{\dag}D_{\mu}U u^{\dag}$.\\
Here the following assumptions have been made:
\begin{enumerate}
\item Before weak gauging, the Lagrangian responsible for EWSB has a $SU\left( 2\right) _{L}\times SU\left( 2\right) _{R}$ global symmetry which is spontaneously broken by the new strong dynamics down to $SU\left(2\right)_{L+R}$. The spontaneous breaking of the global symmetry also leads to the breaking of the standard electroweak gauge symmetry, $SU(2)_L\times U(1)_Y$, down to the electromagnetic $U(1)$.
\item The strong dynamics produces only one triplet of vectors $V_{\mu}^{a}$ belonging to the adjoint representation of the $SU\left(2\right)_{L+R}$ custodial group. These vector states are mass degenerate and their mass is below the cutoff $\Lambda\approx 3\,\text{TeV}$. The parity odd heavy vectors are integrated out. The new $V$ states couple to fermions only via SM gauge interactions.
\end{enumerate}

\section{Gauge model scenario.}
The scattering amplitudes for the processes $W_{L}W_{L}\rightarrow V_{L}V_{L}$ and $W_{L}W_{L}\rightarrow V_{L}V_{T}$ will grow at most as $\frac{s}{v^{2}}$ and the $q\overline{q}\rightarrow VV$ scattering amplitude will go as a constant only when \cite{Barbieri:2010}: 
\begin{equation}
g_{K}=\frac{1}{g_{V}},\qquad f_{V}=2g_{V},\qquad g_{1}=g_{2}=g_{4}=g_{5}=0,\qquad g_{3}=-\frac{1}{4},\qquad g_{6}=\frac{1}{2}
\end{equation}
This choice of parameters defines the Gauge Model scenario, which corresponds to the case in which the new heavy vector states are the gauge vectors arising from a spontaneously broken gauge symmetry $G=SU\left( 2\right) _{L}\otimes SU\left( 2\right) _{R}\otimes SU\left( 2\right) _{C}$ to the diagonal subgroup $SU\left( 2\right) _{L+R+C}$ by non linear sigma model \cite{Barbieri:2010}. The asymptotic behaviour of the $W_{L}W_{L}\rightarrow V_{L}V_{L}$ and $W_{L}W_{L}\rightarrow V_{L}V_{T}$ scattering amplitudes can be improved by the addition of a composite scalar as discussed in \cite{Carcamo:2010}.
\section{Vector Pair production cross sections by Vector Boson Fusion.}
The total cross sections at the LHC for heavy vector pair production via Vector Boson Fusion (VBF) in the gauge and composite model scenarios are respectively shown in Figures 1.a and 1.b. The heavy vector mass is taken to range from $400$~GeV to $800$~GeV. The coupling $G_V$ of the heavy vector to two longitudinal Standard Model gauge bosons is taken to be $G_V=200$~GeV, which is in the allowed region consistent with the unitarity constraint in the elastic channel for longitudinal SM gauge boson scattering \cite{Barbieri:2008}. In the composite model scenario, all the parameters are kept as in the gauge model with the exception of $g_K g_V = 1/\sqrt{2}$. 
These total cross sections have been computed by using the Matrix Element Generator CalcHEP and setting the acceptance cuts $p_T > 30$~GeV and $|\eta|< 5$ for the forward quark jets. These cross sections, which involve all polarizations of the intermediate light bosons, have also a weak dependence on $F_V$, which is set to zero. Since in the gauge model scenario the growth of the $WW\to VV$ scattering amplitude is more under control than in the composite model scenario, the total cross sections for vector pair production via VBF for the gauge model have lower values than in the composite model, as shown in Figures 1.a and 1.b.
\begin{figure}[tbp]
\begin{minipage}[b]{8.2cm}
   \centering
   \includegraphics[width=5.6cm]{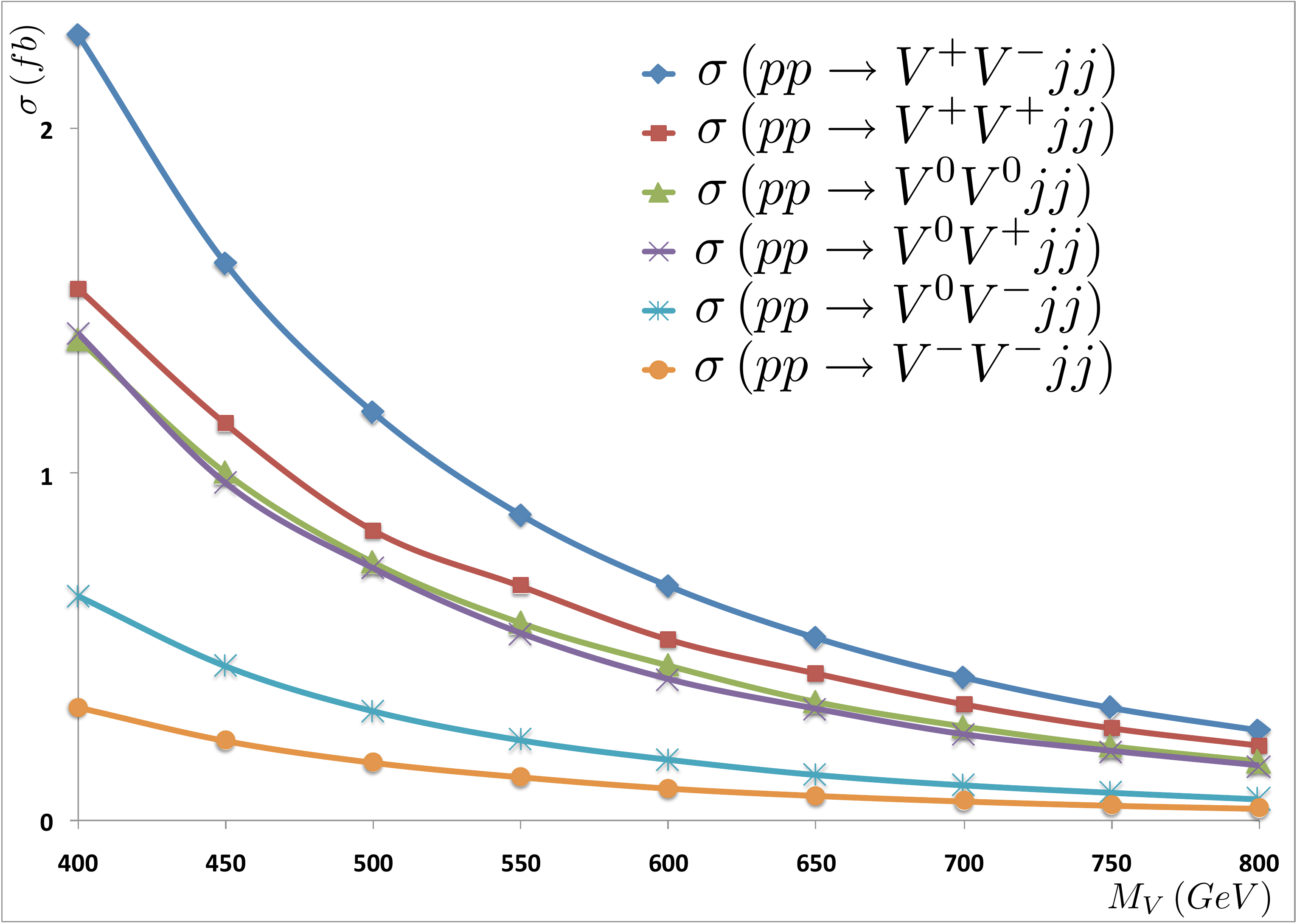}\\\vspace{4mm}\footnotesize{(1.a)}
 \end{minipage}
\begin{minipage}[b]{8.5cm}
  \centering
   \includegraphics[width=5.6cm]{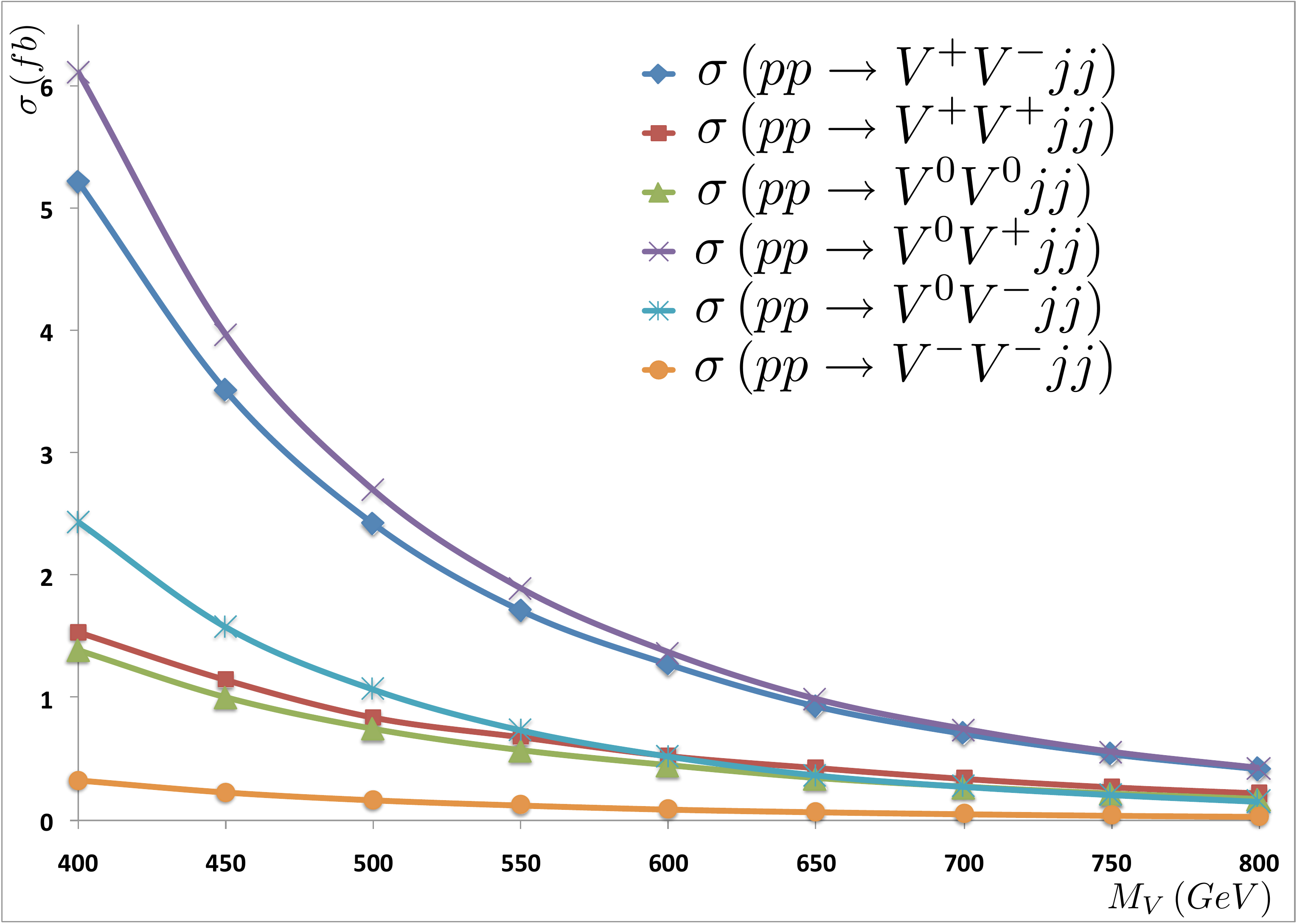}\\\vspace{4mm}\footnotesize{(1.b)}
 \end{minipage}
\caption{Total cross sections for Vector pair production via Vector Boson fusion in a Gauge Model (1.a) and in a Composite Model (1.b)\cite{Barbieri:2010}}
\label{Fig_VBF}
\end{figure}

\section{Pair production cross sections by Drell-Yan annihilation.}
The total cross sections at the LHC for heavy vector pair production via Drell-Yan annihilation as functions of the heavy vector mass in the gauge and composite model scenarios are respectively shown in Figures 2.a and 2.b. The heavy vector mass is taken to range from $400$~GeV to $800$~GeV. As for the vector boson fusion, $G_V=200$~GeV while in the composite model scenario the para-meters are kept as in the gauge model except for $g_K g_V = 1/\sqrt{2}$. Since in the gauge model scenario the Drell-Yan vector pair production amplitudes go as a constant at high energies, the total cross sections in this case are lower than those in the composite model.
\begin{figure}[tbp]
\begin{minipage}[b]{8.2cm}
   \centering
   \includegraphics[width=5.6cm]{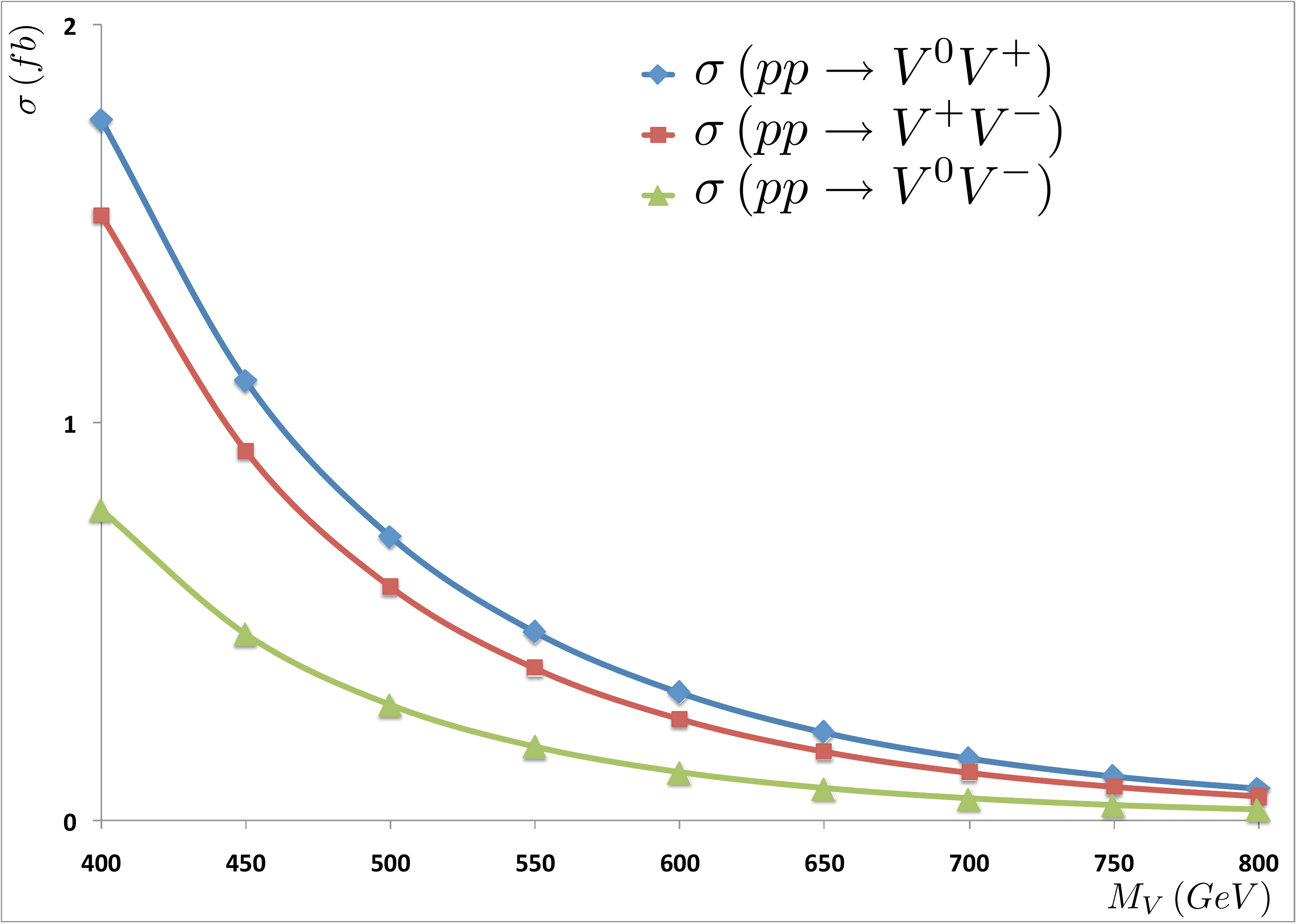}\\\vspace{4mm}\footnotesize{(2.a)}
 \end{minipage}
\begin{minipage}[b]{8.5cm}
  \centering
   \includegraphics[width=5.6cm]{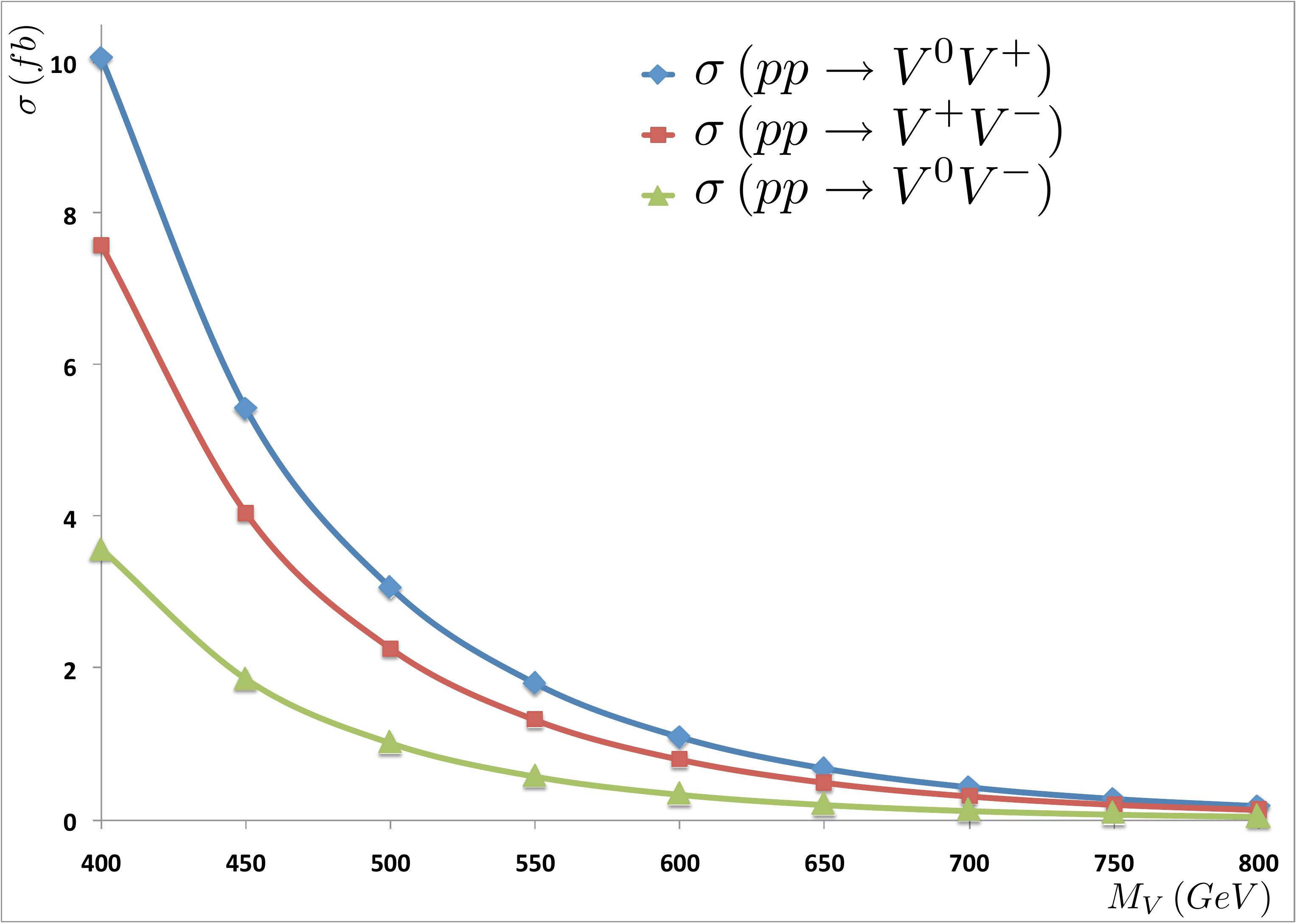}
	\\\vspace{4mm}\footnotesize{(2.b)}
 \end{minipage}
\caption{Total cross sections for Vector pair production via Drell-Yan annihilation in a Gauge Model (2.a) and in a Composite Model (2.b)\cite{Barbieri:2010}}
\label{Fig_DY}
\end{figure}

\section{Same-sign di-lepton and tri-lepton events at LHC.}
Since the heavy vector have dominant decay mode into pair of SM Gauge bosons with branching ratio very close to one, the vector pair production by VBF and DY will lead to 4 SM gauge bosons in the final state. The following table shows the total number of same sign di-lepton and tri-lepton events ($e$ or $\protect\mu$ from $W$ decays) at the LHC for a reference integrated luminosity of $\int\mathcal{L}dt=100~\text{fb}^{-1}$. The heavy vector mass is taken to be $M_V= 500$~GeV.
\begin{table}[ht!]
\begin{center}
{\small 
\begin{tabular}{|c|c|c|}
\hline
& di-leptons & tri-leptons \\ \hline
Vector Boson Fusion (Gauge Model) & 16 & 3 \\ \hline
Drell Yan annihilation (Gauge Model) & 5 & 1 \\ \hline
Vector Boson Fusion (Composite Model) & 28 & 6 \\ \hline
Drell Yan annihilation (Composite Model) & 18 & 4 \\ \hline
\end{tabular}
}
\end{center}
\label{tabcut}
\caption{Number of same sign di-lepton and trilepton events at the LHC}
\end{table}

\section{Conclusions}
In the framework of strongly interacting dynamics for EWSB, composite heavy vector states may exist and the interactions among themselves and with the Standard Model gauge bosons can be described by a $SU(2)_L\times SU(2)_R/SU(2)_{L+R}$ Effective Chiral Lagrangian. The total cross sections at the LHC for the vector pair production by Vector Boson Fusion and Drell-Yan annihilation are of order of few $fb$. The numbers of same sign Dilepton and Trilepton events at the LHC with an integrated luminosity of $100fb^{-1}$ are of order of $10$. Further detailed studies will have to be made to assess the detectability of these processes above the Standard Model backgrounds.
\subsection*{Acknowledgements}
This work was done in collaboration with Riccardo Barbieri, Gennaro Corcella, Riccardo Torre and Enrico Trincherini. It is pleasure to thank all of them for this fruitful work. I also thank the organizers of the XVIII International DIS2010 Workshop, especially Gennaro Corcella for inviting me to present this talk. 


\newpage
\begin{thebibliography}{99}
\bibitem{Barbieri:2008} R. Barbieri, G. Isidori, V. S. Rychkov and E.
Trincherini, Phys. Rev. D \textbf{78} (2008) 036012 [\href{http://arXiv.org/pdf/0911.1942}{arXiv:0911.1942[hep-ph]}].
\bibitem{Barbieri:2010} R.~Barbieri, A.~E.~C\'arcamo Hern\'andez, G.~Corcella, R.~Torre and E.~Trincherini, JHEP {\bf 03} (2010) 068 [\href{http://arXiv.org/pdf/0911.1942}{arXiv:0911.1942[hep-ph]}] 
\bibitem{Carcamo:2010} A.~E.~C\'arcamo Hern\'andez and R.~Torre [\href{http://arXiv.org/pdf/1005.3809}{arXiv:1005.3809[hep-ph]}], accepted for publication in Nuclear Physics B.
\bibitem{Torre:2010}R.~Torre, [\href{http://arXiv.org/pdf/1005.4801}{arXiv:1005.4801[hep-ph]}]
\bibitem{Carcamo:2009}A.~E.~C\'{a}rcamo Hern\'{a}ndez, Proceedings of the First Young Researchers Workshop "Physics Challenges in the LHC Era" 2009, Ed. E. Nardi, Frascati, May 11th and May 14th, 2009 [\href{http://www.lnf.infn.it/sis/frascatiseries/Volume48/volume48.pdf}{www.lnf.infn.it/sis/frascatiseries/Volume48/volume48.pdf}] 
\bibitem{Carcamo:2010ggtoVV}A. E. C\'{a}rcamo Hern\'{a}ndez, [\href{http://arXiv.org/pdf/1008.1039}{arXiv:1008.1039[hep-ph]}]
\bibitem{Carcamo:2010ckm}A. E. C\'{a}rcamo Hern\'{a}ndez and Rakibur Rahman [\href{http://arXiv.org/pdf/1007.0447}{arXiv:1007.0447[hep-ph]}].
\bibitem{Cata:2009iy} O.~Cata, G.~Isidori and J.~F.~Kamenik, 
Nucl.\ Phys.\ B \textbf{822} (2009) 230 [\href{http://arXiv.org/pdf/0905.0490}{arXiv:0905.0490[hep-ph]}].
\bibitem{Pelaez:1996} J.~R.~Pel\'aez, Phys.\ Rev.\ D \textbf{55} (1997) 4193 [\href{http://arxiv.org/abs/hep-ph/9609427}{arXiv:hep-ph/9609427}].

\bibitem {Kaplan:1983} D.~B.~Kaplan and H.~Georgi,
Phys.\ Lett.\ B \textbf{136} (1984) 183.

\bibitem {Chivukula:1993} R.~S.~Chivukula and V.~Koulovassilopoulos,
Phys.\ Lett.\ B \textbf{309}, 371 (1993)
[\href{http://arxiv.org/abs/hep-ph/9304293}{arXiv:hep-ph/9304293}].

\bibitem {Contino:2009}R.~Contino,
[\href{http://arxiv.org/abs/0908.3578}{arXiv:0908.3578 [hep-ph]}].

\bibitem {ggpr}G. F. Giudice, C. Grojean, A. Pomarol and R. Rattazzi, JHEP
0706 (2007) 045
[\href{http://arxiv.org/abs/hep-ph/0703164}{arXiv:hep-ph/0703164}].

\bibitem {Low:2009}I.~Low, R.~Rattazzi and A.~Vichi,
[\href{http://arxiv.org/abs/hep-ph/0907.5413}{arXiv:hep-ph/0907.5413}]

\bibitem{Zerwekh:2010}A. R. Zerwekh, Mod. Phys. Lett. A A25 (2010), 423 [\href{http://arxiv.org/abs/hep-ph/0907.4690}{arXiv:hep-ph/0907.4690}].

\bibitem {Bagger}J.~Bagger~\emph{et al.,} Phys.\ Rev.\ D \textbf{49} (1994) 1246.


\bibitem{Pelaez:1996} J.~R.~Pel\'aez, Phys.\ Rev.\ D \textbf{55} (1997) 4193 [\href{http://arxiv.org/abs/hep-ph/9609427}{arXiv:hep-ph/9609427}].

\bibitem {SekharChivukula:2001} R.~S.~Chivukula, D.~A.~Dicus and H.~J.~He,
Phys.\ Lett.\ B \textbf{525} (2002) 175
[\href{http://arxiv.org/abs/hep-ph/0111016}{arXiv:hep-ph/0111016}].

\bibitem {Csaki:2003}C.~Csaki, C.~Grojean, H.~Murayama, L.~Pilo and
J.~Terning,
Phys.\ Rev.\ D \textbf{69}, 055006 (2004)
[\href{http://arxiv.org/abs/hep-ph/0305237}{arXiv:hep-ph/0305237}].


\bibitem{Zerwekh:2006}A. R. Zerwekh, Eur. Phys. J. C 46 (2006) 791 [\href{http://arxiv.org/abs/hep-ph/0512261}{arXiv:hep-ph/0512261}].



\bibitem {Appelquist:2003}T.~Appelquist and R.~Shrock, Phys. Rev. Lett. 90, 201801 (2003), [\href{http://arxiv.org/abs/hep-ph/0301108}{arXiv:hep-ph/0301108}].
\bibitem{Quigg:2009}C.~Quigg and R.~Shrock, Phys. \ Rev. \ D \textbf{79} :096002 (2009) [\href{http://arxiv.org/abs/hep-ph/0901.3958}{arXiv:hep-ph/0901.3958}].
\bibitem{Han}T. Han, D. L. Rainwater and G. Valencia, Phys. \ Rev. \ D \textbf{68} 015003 (2003) [\href{http://arxiv.org/abs/hep-ph/0301039}{arXiv:hep-ph/0301039}].
\bibitem{Sanino:2008}R. Foadi, M. Jarvinen and F. Sannino, Phys. \ Rev. \ D \textbf{79} (2008) 035010 [\href{http://arxiv.org/abs/hep-ph/0811.3719}{arXiv:hep-ph/0811.3719}].
\bibitem{Barbieri:2008b} R. Barbieri, G. Isidori and D.
Pappadopulo, JHEP {\bf 02} (2009) 029
  [\href{http://arxiv.org/abs/0811.2888}{arXiv:0811.2888 [hep-ph]}].

\bibitem {He:2007ge}H.~J.~He \textit{et al.},
Phys.\ Rev.\ D \textbf{78} (2008) 031701
[\href{http://arXiv.org/pdf/0708.2588}{arXiv:0708.2588[hep-ph]}].


\bibitem {Accomando:2008jh}E.~Accomando, S.~De Curtis, D.~Dominici and
L.~Fedeli,
Phys. Rev. D \textbf{79} (2009) 055020 [\href{http://arXiv.org/abs/hep-ph/0807.5051}{arXiv:hep-ph/0807.5051}]; Nuovo
Cim.\ \textbf{123B} (2008) 809 [\href{http://arXiv.org/abs/hep-ph/0807.2951}{arXiv:hep-ph/0807.2951}].


\bibitem {Belyaev:2008yj}A.~Belyaev, R.~Foadi, M.~T.~Frandsen, M.~Jarvinen,
F.~Sannino and A.~Pukhov,
Phys.\ Rev.\ D \textbf{79} (2009) 035006, [\href{http://arXiv.org/abs/hep-ph/0809.0793}{arXiv:hep-ph/0809.0793}].

\bibitem{Hirn:2007}
  J.~Hirn, A.~Martin and V.~Sanz,
  JHEP {\bf 0805} (2008) 084
 [\href{http://arXiv.org/abs/hep-ph/0712.3783}{arXiv:hep-ph/0712.3783}].
  Phys.\ Rev.\  D {\bf 78} (2008) 075026
  [\href{http://arXiv.org/abs/hep-ph/0807.2465}{arXiv:hep-ph/0807.2465 }].



\bibitem {Ecker:1988te}G.~Ecker, J.~Gasser, A.~Pich and E.~de Rafael,
Nucl.\ Phys.\ B \textbf{321} (1989) 311.

\bibitem {Ecker:1989yg}G.~Ecker, J.~Gasser, A.~Pich and E.~de Rafael,
Phys.\ Lett.\ B \textbf{223} (1989) 425;

\bibitem {coleman}S.~R.~Coleman \textit{et. al.} Phys.\ Rev.\ \textbf{177},
2239, 2247 (1969);
C.G.~Callan, \textit{et. al.}
Phys.\ Rev.\ \textbf{177} (1969) 2247.

\bibitem{Contino:2010t}C.~Contino,
[\href{http://arxiv.org/abs/1005.4269v1}{arXiv:1005.4269v1 [hep-ph]}].

\bibitem {Contino:2010}R.~Contino, C.~Grojean, M.~Moretti, F.~Piccinini and
R.~Rattazzi, [\href{http://arXiv.org/pdf/1002.1011}{arXiv:1002.1011[hep-ph]}]

\bibitem {Casalbuoni:1985}R.~Casalbuoni, S.~De Curtis, D.~Dominici and
R.~Gatto,
Phys.\ Lett.\ B \textbf{155} (1985) 95; Nucl.\ Phys.\ B \textbf{282} (1987) 235.


\bibitem {Nomura}Y.~Nomura,
JHEP \textbf{0311} (2003) 050 [\href{http://arXiv.org/abs/hep-ph/0309189}{arXiv:hep-ph/0309189}].



\bibitem {Barbieri:2003pr}R.~Barbieri, A.~Pomarol and R.~Rattazzi,
Phys.\ Lett.\ B \textbf{591} (2004) 141  [\href{http://arxiv.org/abs/hep-ph/0310285}{arXiv:hep-ph/0310285}].

\bibitem {Foadi:2003xa}R.~Foadi, S.~Gopalakrishna and C.~Schmidt,
JHEP \textbf{0403} (2004) 042 [\href{http://arxiv.org/abs/hep-ph/0312324}{arXiv:hep-ph/0312324}].


\bibitem {Georgi:2004iy}H.~Georgi,
Phys.\ Rev.\ D \textbf{71} (2005) 015016 [\href{http://arxiv.org/abs/hep-ph/0408067}{arXiv:hep-ph/0408067}].

\bibitem {SekharChivukula:2008mj} R.~S.~Chivukula, H.~J.~He, M.~Kurachi,
E.~H.~Simmons and M.~Tanabashi,
Phys.\ Rev.\ D \textbf{78}, 095003 (2008), [\href{http://arxiv.org/abs/hep-ph/0808.1682}{arXiv:hep-ph/0808.1682}].

\bibitem {calchep}A. Pukhov, A. Belyaev and N. Christensen,
\href{http://theory.sinp.msu.ru/~pukhov/calchep.html}{http://theory.sinp.msu.ru/$\sim
$pukhov/calchep.html}.


\bibitem{Foadi:2008} R.~Foadi, M.~J\"arvinen and F.~Sannino, Phys.\ Rev.\ D \textbf{79} (2008) 035010 [\href{http://arXiv.org/abs/0811.3719}{arXiv:0811.3719 [hep-ph]}].

\bibitem {feynrules}N. Christensen, C. Duhr and B. Fucks,
\href{http://feynrules.phys.ucl.ac.be/}{http://feynrules.phys.ucl.ac.be/}


\bibitem {Chivukula:2003}R.~S.~Chivukula, D.~A.~Dicus, H.~J.~He and
S.~Nandi,
Phys.\ Lett.\ B \textbf{562} (2003) 109 [\href{http://arXiv.org/abs/hep-ph/030226}{arXiv:hep-ph/030226}].


\bibitem {Birkedal:2005yg}A.~Birkedal, K.~T.~Matchev and M.~Perelstein,
\textit{In the Proceedings of 2005 International Linear Collider Workshop
(LCWS 2005), Stanford, California, 18-22 Mar 2005, pp 0314}
[\href{http://arXiv.org/pdf/0508185}{arXiv:0508185[hep-ph]}]

\bibitem{Kaplan:1991dc}
  D.~B.~Kaplan,
  Nucl.\ Phys.\  B {\bf 365} (1991) 259.

\bibitem{Grojean}C.~Grojean,
[\href{http://arxiv.org/abs/0910.4976v1}{arXiv:0910.4976v1 [hep-ph]}].




\bibitem{Isidori}G.~Isidori,
[\href{http://arxiv.org/abs/0911.3219v1}{arXiv:0911.3219v1 [hep-ph]}].
\bibitem{Manohar} A. V. Manohar, [\href{http://arxiv.org/abs/hep-ph/9606222v1}{arXiv:hep-ph/9606222v1}].
\bibitem{Pich} A. Pich, [\href{http://arxiv.org/abs/hep-ph/9806303v1}{arXiv:hep-ph/9806303v1}].
\bibitem{Hirn:2004}
  J.~Hirn and J.~Stern,
  Eur.\ Phys.\ J.\  C {\bf 34} (2004) 447 [\href{http://arXiv.org/abs/hep-ph/0401032}{arXiv:hep-ph/0401032}].

\bibitem {Pallante:1992qe}E.~Pallante and R.~Petronzio,
Nucl.\ Phys.\ B \textbf{396} (1993) 205.



\bibitem {Borasoy:1995ds}B.~Borasoy and U.~G.~Meissner,
Int.\ J.\ Mod.\ Phys.\ A \textbf{11} (1996) 5183  [\href{http://arxiv.org/abs/hep-ph/9511320}{arXiv:hep-ph/9511320}].



\bibitem {Harada:2003jx}M.~Harada and K.~Yamawaki,
Phys.\ Rept.\ \textbf{381} (2003) 1  [\href{http://arxiv.org/abs/hep-ph/0302103}{arXiv:hep-ph/0302103}].



\bibitem {Bijnens:1995ii}J.~Bijnens and E.~Pallante,
Mod.\ Phys.\ Lett.\ A \textbf{11} (1996) 1069 [\href{http://arxiv.org/abs/hep-ph/9510338}{arXiv:hep-ph/9510338}].



\bibitem {Cirigliano:2006hb}V.~Cirigliano, G.~Ecker, M.~Eidemuller, R.~Kaiser,
A.~Pich and J.~Portoles,
Nucl.\ Phys.\ B \textbf{753} (2006) 139 [\href{http://arxiv.org/abs/hep-ph/0603205}{arXiv:hep-ph/0603205}].



\bibitem {Kampf:2006yf}K.~Kampf, J.~Novotny and J.~Trnka,
Eur.\ Phys.\ J.\ C \textbf{50} (2007) 385 [\href{http://arXiv.org/pdf/0608051}{arXiv:[hep-ph/0608051]}].

\end{thebibliography}
\end{document}